\documentclass[aip,preprint]{revtex4-1}
\usepackage{graphicx}
\usepackage{dcolumn}
\usepackage{bm}
\usepackage{color}
\usepackage{tabularx}
\usepackage{array}
\usepackage{amsmath}
\usepackage{amsfonts}
\usepackage{pbox}
\usepackage{stmaryrd}  
\usepackage{floatrow}
\usepackage{sidecap}

\begin{document}

\title{Magnon-Photon Coupling in an Opto-Electro-Magnonic Oscillator}

\author{Yuzan Xiong}
\thanks{Equal Contribution.}
\affiliation{Department of Physics and Astronomy, University of North Carolina at Chapel Hill, Chapel Hill, NC 27599}

\author{Jayakrishnan M. P. Nair}
\thanks{Equal Contribution.}
\affiliation{Department of Physics,Boston College,140 Commonwealth Avenue,Chestnut Hill, MA 02467}

\author{Andrew Christy}
\affiliation{Department of Physics and Astronomy, University of North Carolina at Chapel Hill, Chapel Hill, NC 27599}
\affiliation{Department of Chemistry, University of North Carolina at Chapel Hill, Chapel Hill, NC 27599}

\author{James F. Cahoon}
\affiliation{Department of Chemistry, University of North Carolina at Chapel Hill, Chapel Hill, NC 27599}

\author{Amin Pishehvar}
\affiliation{Department of Electrical and Computer Engineering, Northeastern University, Boston, MA 02115, USA}

\author{Xufeng Zhang}
\affiliation{Department of Electrical and Computer Engineering, Northeastern University, Boston, MA 02115, USA}

\author{Benedetta Flebus}
\thanks{flebus@bc.edu} 
\affiliation{Department of Physics,Boston College,140 Commonwealth Avenue,Chestnut Hill, MA 02467}

\author{Wei Zhang}
\thanks{zhwei@unc.edu}
\affiliation{Department of Physics and Astronomy, University of North Carolina at Chapel Hill, Chapel Hill, NC 27599}

\date{\today}

\begin{abstract}
\centerline{\textbf{Abstract}}

The opto-electronic oscillators (OEOs) hosting self-sustained oscillations by a time-delayed mechanism are of particular interest in long-haul signal transmission and processing. On the other hand, owing to their unique tunability and compatibility, magnons - as elementary excitations of spin waves - are advantageous carriers for coherent signal transduction across different platforms. In this work, we integrated an opto-electronic oscillator with a magnonic oscillator consisting of a microwave waveguide and a yttrium iron garnet sphere. \textcolor{black}{We find that, in the presence of the magnetic sphere, the oscillator power spectrum exhibits sidebands flanking the fundamental OEO modes. The measured waveguide transmission reveals anti-crossing gaps, a hallmark of the coupling between the opto-electronic oscillator modes and the Walker modes of the sphere. Experimental results are well reproduced by a coupled-mode theory that accounts for nonlinear magnetostrictive interactions mediated by  the magnetic sphere}. Leveraging the \textcolor{black}{advanced} fiber-optic technologies in opto-electronics, this work lays out a new, hybrid platform for investigating long-distance coupling and nonlinearity in coherent magnonic phenomena.

\end{abstract}

\maketitle

\section{Introduction}

Hybrid magnonics \cite{nakamura_apex2019,hu_ssp2018,bhoi_ssp2019,tqe_2021,jap_2021} have witnessed immense developments in the past decades, with successful demonstrations in the coherent coupling between magnons and other fundamental excitations including microwave, light, phonon, and qubit \cite{nakamura_science2015,haigh_prb2019,yili_prl2019,luqiao_prl2019,huebl_prl2013,bai_prl2015,xufeng_prl2014}. In addition to the continuous striving towards high coupling strength (high cooperativity) as well as versatile control knobs for hybridization, two other fundamental constituents in hybrid magnonics, namely, the `nonlinearity' \cite{nonlinearity_yuan2023,bauer_prb2020} and `long-distance coupling' \cite{prl_rao2023} have been  receiving increasing attention. These two subjects were nurtured by recent experimental discoveries, namely, the `gain-driven polariton (GDP)' \cite{gdp}, in which the light-matter interaction is activated and sustained by sourcing from a dc gain (amplification) and the `pump-induced mode (PIM)' \cite{pim}, where a collective excitation of non-saturated spins are pumped (driven forcefully away) from its equilibrium, in magnon-photon coupled systems. 

\textcolor{black}{Thus far, these phenomena have only been explored in the \textcolor{black}{'local, non-distributed' fashion} dictated by the close proximity required for realizing strong magnon-photon excitations.  A natural candidate for overcoming this constraint is the opto-electronic oscillator} \cite{larger_rmp2019}. \textcolor{black}{Some of its remarkable attributes include}: (i) high quality(Q)-factor microwave photon mode in the GHz regime, \textcolor{black}{which} is generated and sustained by a dc-gain driven mechanism; (ii) the characteristic frequency and sidebands are governed by the generic time-delay and other phase-delay components; (iii) long-haul signal transmission with minimum insertion loss over kms leveraging fiber optics and convenient Optical(O)-Electrical(E) conversion and vice versa (E-O) by using electro-optic modulators (EOMs). This system, combined with a magnonic time-delay component, e.g., a Y$_3$Fe$_5$O$_{12}$ (YIG) delayline in a magnonic-opto-electronic oscillator \cite{moeo}, has been used and proposed in a few relevant applications, including but not limited to, neuromorphic computing \cite{ustinov_magnlett2015,vitko_magnlett2018,watt1,watt2,watt3}. However, the potential of such OEO mode in  magnon-based hybrid systems is yet vastly unexplored.

\begin{figure}[b]
 \centering
 \includegraphics[width=5.6 in]{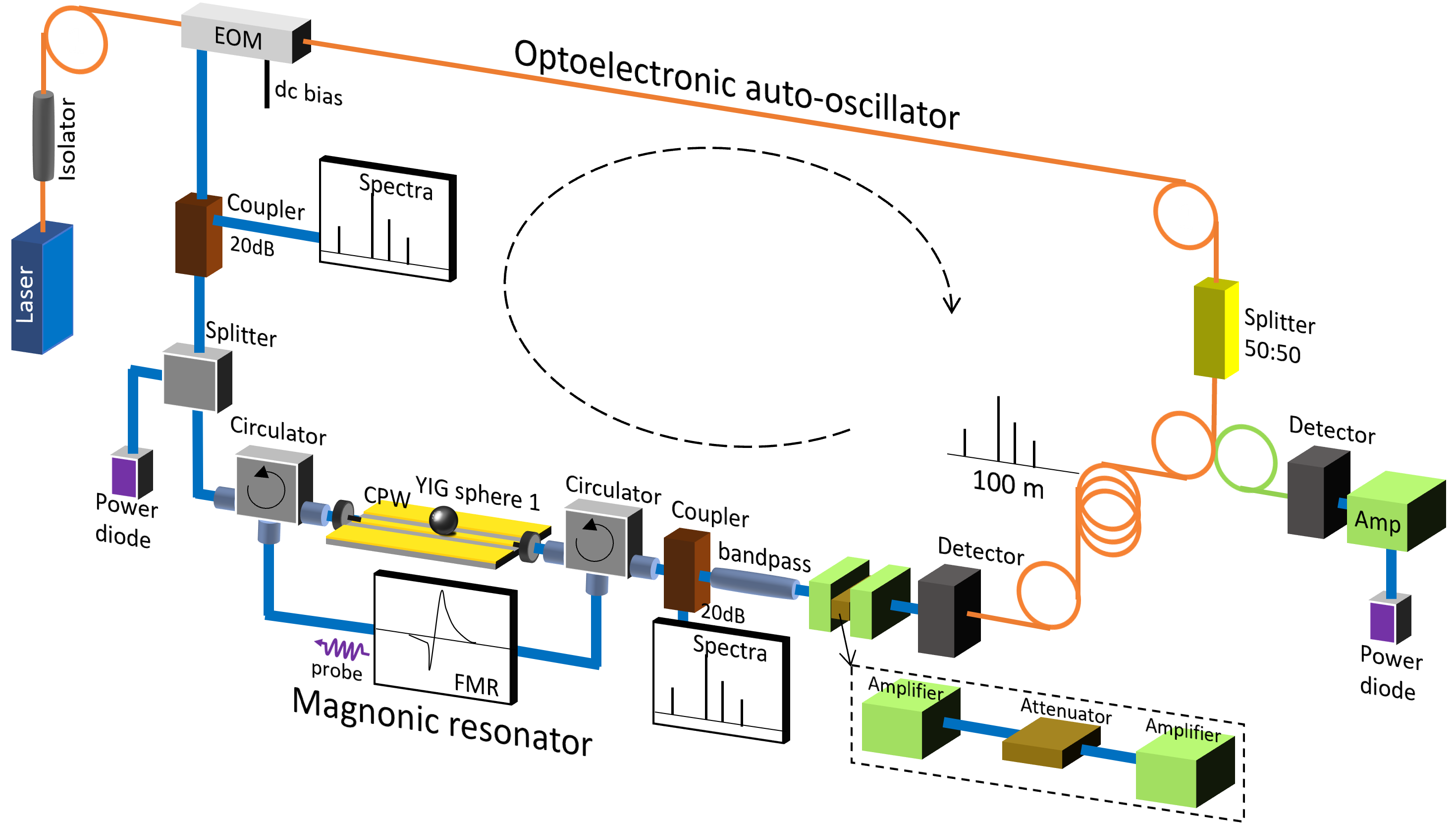}
 \caption{Schematic illustration of the setup consisting of the (main loop) optoelectronic, magnonic (YIG sphere), and fiber delay parts. Bandpassing functions are given by the enclosed circulator pair and dedicated bandpass filters while other rf components are rated broadband. The fiber splitter splits half of the optical power to a microwave diode for system power monitoring. In between the two circulators, a ferromagnetic resonance probe (sub-loop) is installed that probes the YIG magnonic characteristics at a much lower rf power compared to the main loop (negligible disturbance). }
 \label{fig1}
\end{figure}

Similar to the GDP photon mode, the OEO photon mode sources energy from the loop's gain, and can travel over long distances to support remote magnon-photon coupling beyond the near fields. In this sense, it is therefore interesting to integrate such OEO photon mode with another magnon mode from a magnonic resonator, e.g., YIG, and study their coherent coupling and hybridization. Such a unique system will also allow to investigate how the self-sustained OEO photon mode couples to a magnonic system in the same loop, and how the energy transforms and dissipates between respective photon and magnon modes upon changing the various external control parameters. 

In this work, we \textcolor{black}{develop a hybrid architecture consisting of an OEO and a YIG sphere integrated on top of a microwave waveguide, as illustrated in Fig. \ref{fig1}. We observe that, in the presence of the YIG sphere, the power spectrum of the OEO exhibits sidebands that can be explained by invoking  nonlinear magnetostrictive interactions between the magnon and phonon modes of the magnetic sphere. Our interpretation is substantiated by numerical simulations of matching phonon modes with frequencies equal to the separation between the fundamental OEO frequency and its sidebands. The  avoided level crossings observed in the transmission data can be ascribed to the hybridization between the OEO modes and Walker modes of the YIG. We find that the level splitting at the fundamental OEO frequency is significantly larger than the one observed in correspondence of the sidebands. This disparity owes its origin to the emergence of a PIM at the fundamental OEO frequency and its intensity-dependent coupling with the photon mode. A coupled-mode theory in conjunction with input-output analysis of the probe transmission is in excellent agreement with the observed transmission data.} 

\section{Results}
\subsection{Setup Construction}

As illustrated in Fig. \ref{fig1}, an intensity-stabilized fiber laser (1550 nm wavelength, power outpout ~17 dBm from Optilab LLC) sources a continuous-wave (CW) light signal that first passes through an isolator, then an electro-optic modulator (EOM) that is biased at the appropriate dc level (0.8 - 1.5 V in our study). The signal is then divided into two branches using a 50:50 fiber light splitter. The right-hand branch can be used as a reference branch or a monitor branch, where the light signal converts to microwave by an ultrafast diode and then amplifies to a power diode for monitoring the optical power of the whole system. 

The main branch (left) is one that entails the magnon resonator: after the light splitting, the signal passes through a fiber delayline, then converts to microwave via an ultrafast diode \textcolor{black}{(Detector)}, and an optional bandpass filter, then amplifies by an \textcolor{black}{``Amplifier-Attenuator-Amplifier"} combo before reaching the magnon resonator. The \textcolor{black}{``Amplifier-Attenuator-Amplifier"} represents a tunable attenuator sandwiched between two fixed-gain amplifiers (+30 \textcolor{black}{dB}) \textcolor{black}{and effectively serves as a ``tunable amplifier",} thus allowing the power level of the signal to be tuned before returning to the EOM (at the rf port). 

The magnon resonator consists of a YIG sphere with a nominal radius $1.0$ mm placed on top of the signal line of a coplanar waveguide (CPW), and the whole structure is placed between a pair of circulators, to introduce an additional, reverse-circulating "probe loop" that measures the ferromagnetic resonance (FMR) of the magnon resonator. An external magnetic field, $H$, is applied to establish the magnon resonances and is set parallel to the CPW's stripline. The probe loop microwave power is kept rather small ($<-$15 dBm) and causes no disturbance to the main loop which typically flows a microwave power $>$ 10 dBm. Once the signal leaves the magnon resonator, it is directed back to the EOM (rf port). Additional microwave splitters and/or couplers are inserted as needed for sampling and monitor the signal power and spectra at desirable positions along the loop. 

\begin{figure}[htb]
 \centering
 \includegraphics[width=5.6 in]{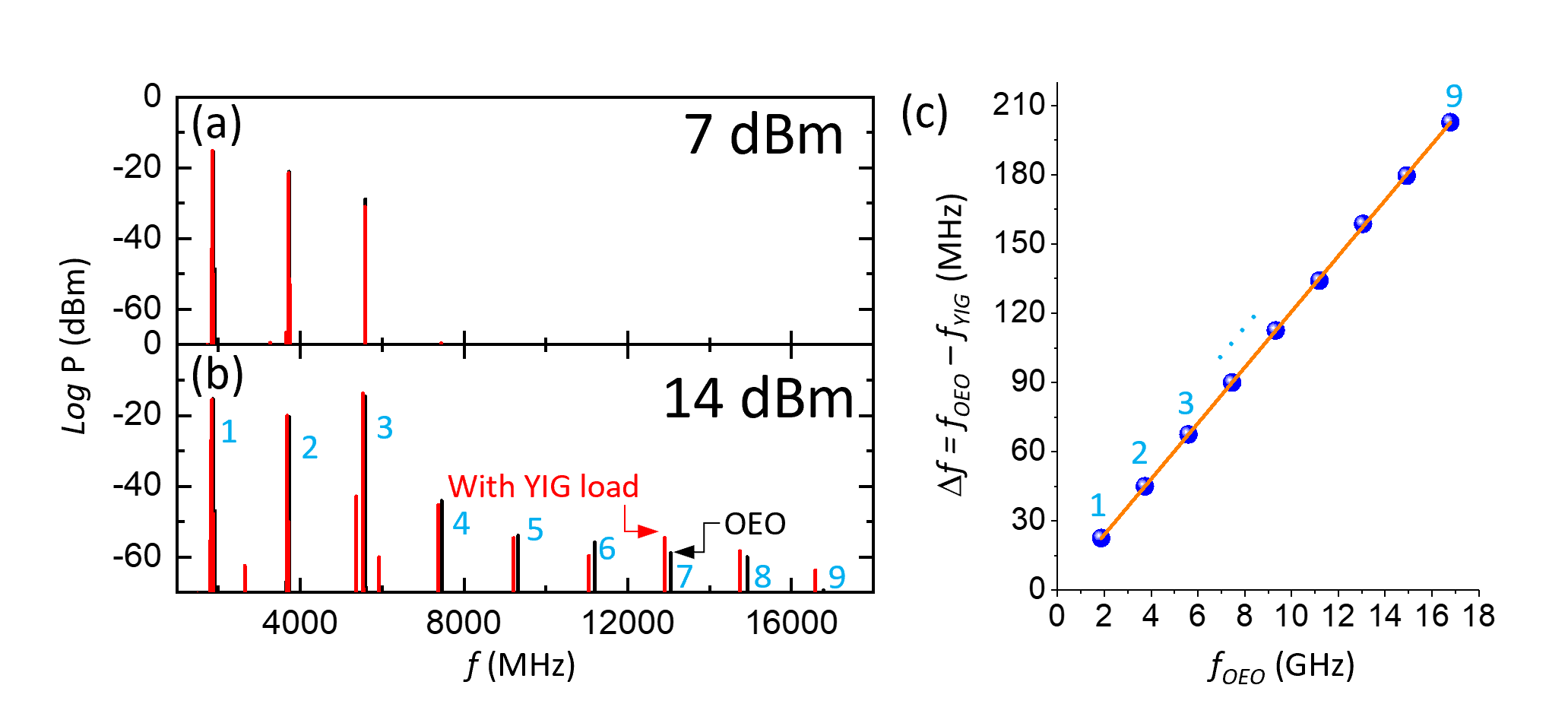}
 \caption{OEO eigen-modes (using a short fiber delayline of 100-m) under two different loop power levels, (a) low power: 7 dBm, and (b) high power: 14 dBm, \textcolor{black}{measured by the rf power diode in the microwave branch.} Power spectra with and without the YIG magnonic load are shown. \textcolor{black}{At 7 dBm loop power, the first 3 eigen-modes are observed. At 14 dBm loop power, a total of 9 discernible eigen-modes are observed, labeled 1 to 9. } (c) The extra magnonic-load induces a frequency shift $\Delta f = f_{OEO}-f_{YIG}$, corresponding to a small additional time delay, which is linearly proportional to the \textcolor{black}{OEO harmonics, $f_{OEO}$ (mode index)}. }
 \label{fig2}
\end{figure}

\subsection{OEO Characteristics and YIG loading}\label{sec3}

Upon a certain \textcolor{black}{microwave} threshold power, self-sustained oscillation sharply establishes in the loop, which we define as the \textcolor{black}{unity-gain} (0 dB), to indicate the onset of the auto-oscillation. A series of eigen-modes in the form of sharp peaks can be observed in the power spectra. \textcolor{black}{These eigen-modes arise because their frequencies satisfy the oscillation condition, which is collectively determined by the loop gain (unity), loop phase ($2\pi N$ with $N$ being an integer), and other active components in the loop, particularly the amplifier gain profile and the attenuator loss profile. \cite{larger_rmp2019}} As shown in Fig. \ref{fig2}(a), for the generic OEO without loading the magnon resonator, the loop power (\textcolor{black}{as measured from the rf power diode in the microwave branch}) to establish and maintain a robust self-oscillation is $\sim 7$ dBm, and 3 eigen-modes can be observed simultaneously. When the power increases, more eigen-modes start to emerge; for example, in Fig. \ref{fig2}(b), up to 9 pronounced eigen-modes were observed when the loop power is tuned to 14 dBm. The position of the modes shift sensitively by the length of the inserted fiber delayline (5, 100, 1000 m are used in this work) as well as the phase shifter, but depends weakly on the loop power (gain). For example, at 14 dBm power and a fiber length of 100 m, we observe a set of fundamental eigen-modes at 1864.4, 3729.6, 5594.8, 7459.4...MHz. These fundamental modes slightly shift to the lower frequency once the magnon resonator is inserted. \textcolor{black}{The shift in frequency, $\Delta f = f_{OEO}-f_{YIG}$ reflects an effective path-length change (increase) by introducing the YIG magnon resonator,} as shown in Fig. \ref{fig2}(c). \textcolor{black}{Linear fitting of $\Delta f$ against the OEO harmonics, $f_{OEO}$, in Fig. \ref{fig2}(c), yields a slope of 0.012, corresponding to a $\sim 1\%$ of path increase after loading the YIG magnon resonator. }

\subsection{Magnon-induced modulation of OEO power and spectra}

\begin{figure}[htb]
 \centering
 \includegraphics[width=6.8 in]{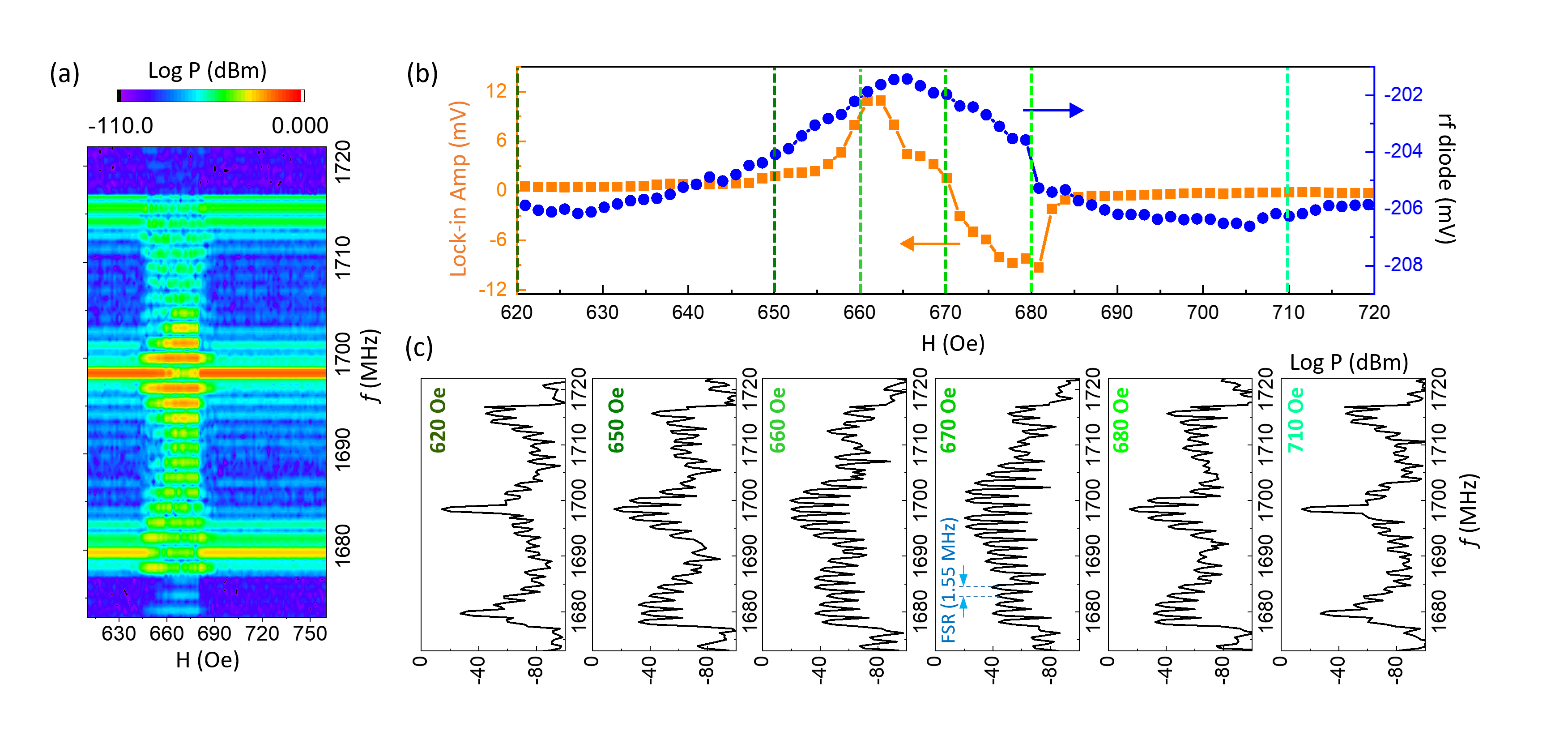}
 \caption{(a) Contour plot of the power spectrum scanned near the FMR regime of a magnon mode driven by the $1^{st}$ OEO photon mode. (b) The FMR signal (probed by the sub-loop at low power) and the power diode signal (main loop) versus the external magnetic field. The power diode measures primarily the power of the OEO main mode, which is at least 20dB higher than all other sidebands at off-resonance. (c) Power spectrum snapshots at selective magnetic fields. Around the magnetic resonance, the OEO's main photon mode is attenuated, whose energy then smears out and distributes to the continuum. \textcolor{black}{The dense oscillations correspond to the free-spectral-range (FSR) of high-order OEO modes, which is 1.55 MHz (645 ns delay) with a 100 m fiber.}}
 \label{fig3}
\end{figure}

We first turn our attention to the 1$^{st}$ fundamental mode under the excitation of the magnon resonator. As an example, we use a fiber of 100 m in length and an additional phase shifter tuned to the optimized phase angle. The main mode resonates at \textcolor{black}{$\omega_1=1698.4$ MHz}, with two pronounced sidebands at 1679.8 and 1715.7 MHz, shown in Fig.~\ref{fig3}. \textcolor{black}{These pronounced sidebands may be attributed to the non-flat gain profile. Near the main oscillation there may exist other frequencies also satisfying the unity gain condition. So the main eigen-modes, although being the most pronounced, may not be the only series that can be supported by the loop. The nearby series can be also prompted due to the small dispersion of the loop gain.} By sweeping the external magnetic field, this fundamental mode excites a YIG FMR at around 665 Oe, inducing a disturbance to the local power spectra near the FMR regime (from 650 to 680 Oe). At the same time, the probe-detected FMR signal and the microwave diode signal (monitoring the loop's optical power) is measured concurrently. Due to the use of a negative video-out diode, an rf power increase corresponds to a diode readout decrease (more negative). As shown in Fig. \ref{fig3}(b), when the FMR is excited as evidenced by the FMR probe \textcolor{black}{(orange trace)}, the optical power of the loop is also correspondingly reduced as indicated by the change (increase) of the diode signal (blue trace). 

\textcolor{black}{Different from using a conventional rf source, the property of the system to monitor the magnon resonator by the OEO loop power is a unique and potentially useful feature of the present dc-gain-driven system. It is likely caused by the similar timescale of the loop recovery to the YIG's relaxation}. The YIG sphere we used has a damping of \textcolor{black}{$\alpha = 1 \times 10^{-4}$} and absorbs/dissipates energy to the OEO's photon mode in the scale of ns; while the OEO relaxation is determined from its delay time, $\tau_d$, can be estimated: $\tau_d = \tau_p +\tau_m$, where $\tau_p$ and $\tau_m$ are the photonic and microwave (including magnonic part) delays, respectively. For long fibers, the time is driven primarily by the photonic delay, i.e., the fiber delay $\tau_p = nL/c$, where $L$ is the fiber length and $n$ the refractive index. The Corning SMF-28 fiber used in our system has a core index of refraction as 1.4682 at 1550 nm, so the speed of light can be estimated as $v = c/n $ 2.04$\times 10^8$ m/s. Therefore, the delay is on the order of $\sim 500$ ns just from the photonic part. \textcolor{black}{A more accurate determination of the total delay can be obtained by analyzing the free-spectral-range (FSR).} The spectrum evolution upon sweeping the magnetic field near the FMR regime is shown in Fig. \ref{fig3}(c) by series of snapshots of the spectrum. As the magnetic field is tuned from 660 Oe to 670 Oe, the amplitudes of the main resonance and its two sidebands gradually reduce, while a denser oscillation pattern shows up in the spectrum, \textcolor{black}{which has an FSR of ~1.55 MHz corresponding to a total time delay of $\sim$ 645 ns for the whole loop with a 100 m fiber and all other components.}  

\subsection{\textcolor{black}{The emergence of photonic sidebands and magnon-photon coupling}}

\begin{figure}[htb]
 \centering
 \includegraphics[width=7 in]{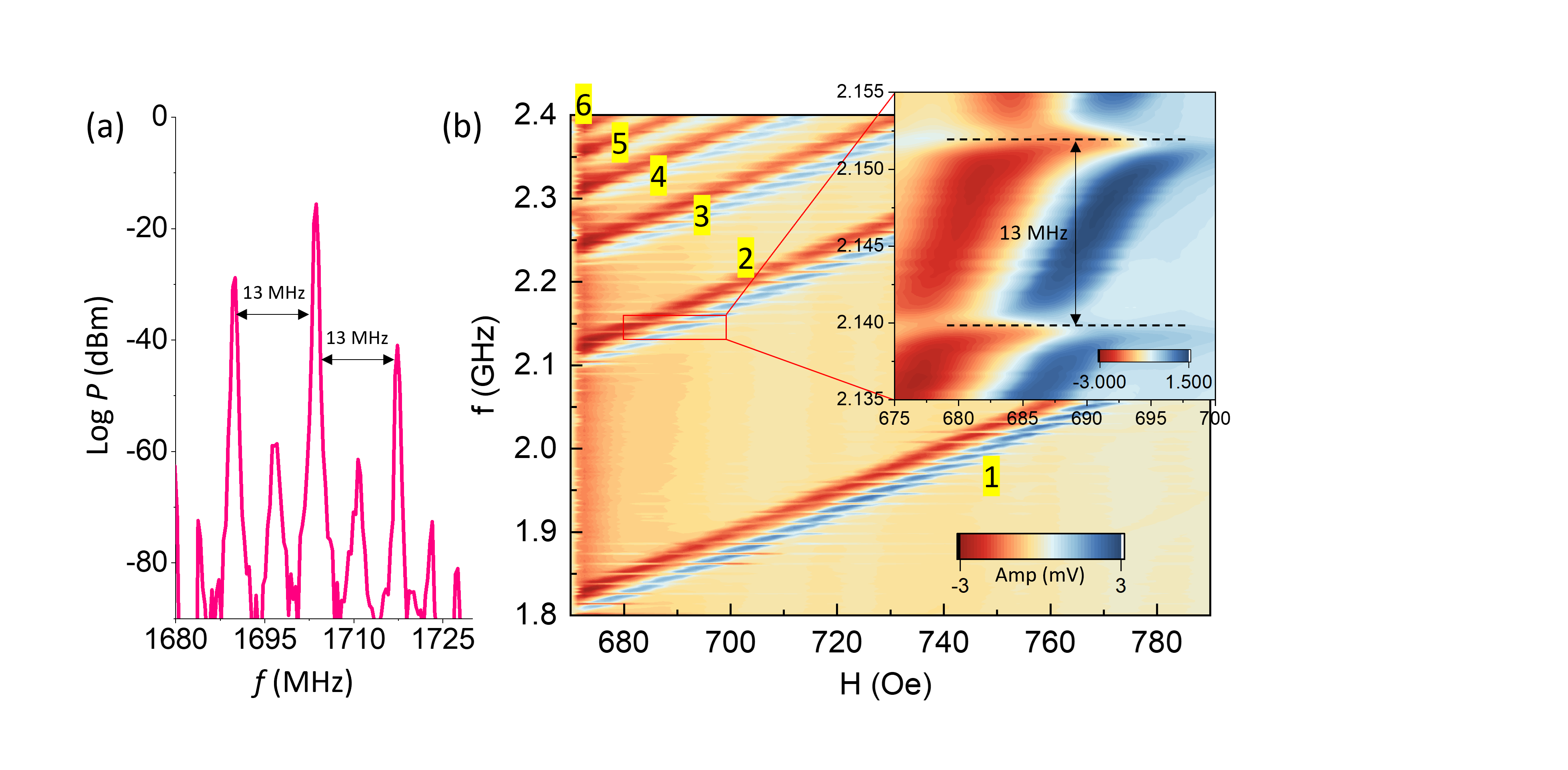}
 \caption{(a) Power spectrum at zero magnetic field showing the main OEO mode and the two pronounced sidebands at a fiber length of 5 m. The left and right sidebands are 13 MHz away from the central band. (b) The scanned [$f,H$] dispersion near the OEO mode by using the probe sub-loop. A series of anticrossing gaps are observed by the hybridization between the mixed photon mode series and the YIG distinct magnetostatic (Walker) modes. }
 \label{fig4}
\end{figure}

\textcolor{black}{Employing a probe sub-loop}, we scanned the dispersion of the magnon resonator near the 1$^{st}$ OEO main mode which resonates at around $H=670$ Oe, \textcolor{black}{as shown in} Fig.~\ref{fig3}. A fiber delay of 5 m is used to mitigate the sensitivity of the spectra to the magnetic field while still observing the needed characteristics. The two sidebands are $\approx 13$ MHz apart from the central band, and all other harmonics are more than 40dB lower in \textcolor{black}{power, as shown in} Fig. \ref{fig4}(a). Under the strong OEO photon mode (akin to a distinct rf signal source), excitation of different magnetostatic modes at different magnetic fields are anticipated. As shown in Fig.\ref{fig4}(b), a total of six magnetostatic modes (i.e., Walker modes / Whisper-gallery modes)\cite{pim} were detected, which are launched near the 1$^{st}$ main OEO mode. 

\textcolor{black}{We can rationalize the emergence of sidebands as a consequence of the nonlinear magnetostrictive interaction between a YIG phonon mode of frequency $\Omega\approx 13$ MHz and the YIG spin waves, which in turn couple to OEO modes.  This interaction, which owes its origin to the deformation of the YIG due to the varying magnetization inside the magnetic sphere \cite{you_prb2016},  leads to the generation of newer frequencies through scattering, e.g., the Stokes and anti-Stokes frequencies of the form $\omega_1-\Omega$, $\omega_1+\Omega$ around the first fundamental frequency $\omega_1$, as shown in Fig. \ref{fig4}(a), and their higher orders. Owing to the dispersion in the loop gain, these modes become self-sustained and can couple with the magnon modes via dipole-dipole interaction, leading to the anti-crossings observed in Fig. \ref{fig4}(b)}. 

\begin{figure}
 \centering
 \includegraphics[width=4.5 in]{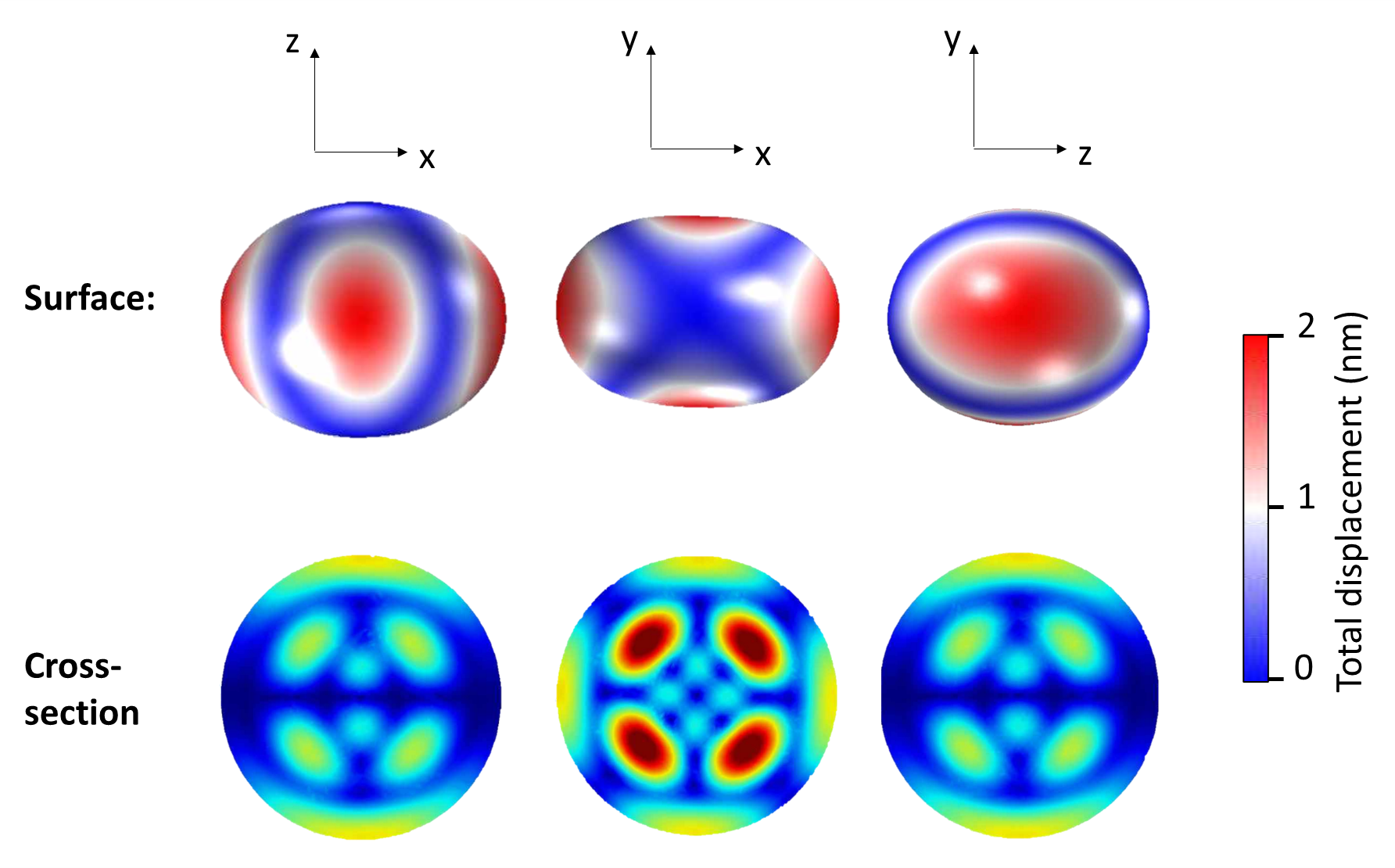}
 \caption{Simulated total displacement of the phonon mode at 13 MHz on a 1-mm-diameter YIG sphere.}
 \label{fig6.5}
\end{figure}

\textcolor{black}{Our COMSOL simulations confirm the existence of a phonon S$_{3,2,2}$ mode at 13 MHz for the 1-mm-diameter YIG sphere, as shown in Fig.~\ref{fig6.5}.  The phonon S$_{3,2,2}$ mode falls into the same symmetry class of the S$_{1,2,2}$ mode discussed in Ref.~[\onlinecite{zhang_sciadv2016}] and thus also couples efficiently with the magnon mode. On the other hand,  its large radial mode order (3) leads to reduced displacement at the sphere surface, which minimizes the contact loss to this phonon mode and accordingly lower self-oscillation threshold.} Hybridization between the magnetostatic and phonon modes were observed in the dispersion, i.e., as a series of anti-crossing gaps that become apparent by a fine-scanned window in the inset of Fig. \ref{fig4}(b). \textcolor{black}{The coupling strength between the photonic sidebands and magnon modes can be extracted from the anti-crossing gap as} \textcolor{black}{ $g_{ma}$$\approx 1.1$ MHz.}    

\section{Discussion}
\subsection{Magnon-photon coupling with the OEO modes}

Due to the relatively low frequency of the first OEO eigen-mode that lays beyond the magnon manifold, the higher order magnetostatic (Walker) modes of YIG cannot be effectively excited except for the main FMR in the probe sub-loop. Therefore, we shift our focus to the $2^{nd}$ OEO eigen-mode at higher frequency, which resonates at \textcolor{black}{$\omega_2=2\omega_1\approx 3778.1$ MHz}. By choosing the proper set of circulators and bandpass filters, we can pick out the $2^{nd}$ OEO mode while suppressing all other eigen-modes and sidebands. 

\begin{figure}[htb]
 \centering
 \includegraphics[width=5 in]{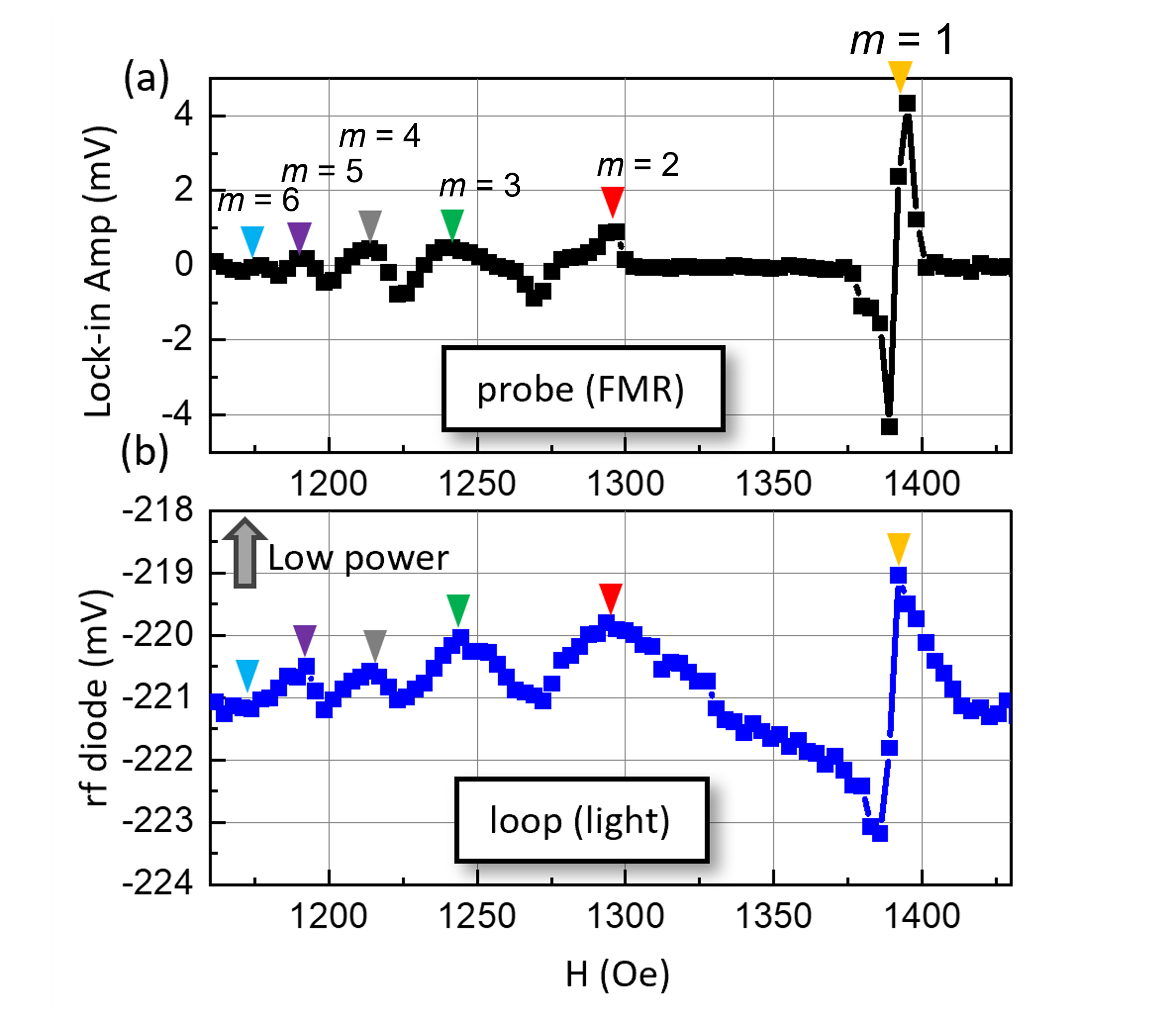}
 \caption{(a) The magnon signal (probed by the sub-loop) and (b) the power diode signal (main loop) versus the external magnetic field. A total of six magnetostatic modes ($m=1,2,3,4,5,6$) of YIG can be observed due to the OEO pump. }
 \label{fig5}
\end{figure}

In the sub-loop, the magnon modes are probed by the transmission of a fixed-frequency rf tone (using the lock-in amplifier) while the bias magnetic field is scanned. Near this 2$^{nd}$ OEO eigen-mode, all the six magnetostatic modes of YIG can be concurrently observed, which is evidenced from both the magnon spectra using the probe sub-loop, in Fig. \ref{fig5}(a), as well as the power diode from the main loop, in Fig. \ref{fig5}(b). These modes are identified as the magnetostatic modes with mode order ($m,m,0$), where $m=1,2,3,...$ is the angular mode number, and the fundamental mode ($m=1$) corresponds to the ferromagnetic resonance (FMR) mode. \textcolor{black}{These modes are circularly polarized. In the direction perpendicular to the bias field, their precession amplitude scales as $\rho^{m-1}$ and precession phase is $(m-1)\phi$. The $m > 1$ modes are non-uniform modes, and they can be excited efficiently when the CPW width is comparable to the YIG sphere radius, which is the present case in our magnonic oscillator. }  

\begin{figure}[htb]
 \centering
 \includegraphics[width=6 in]{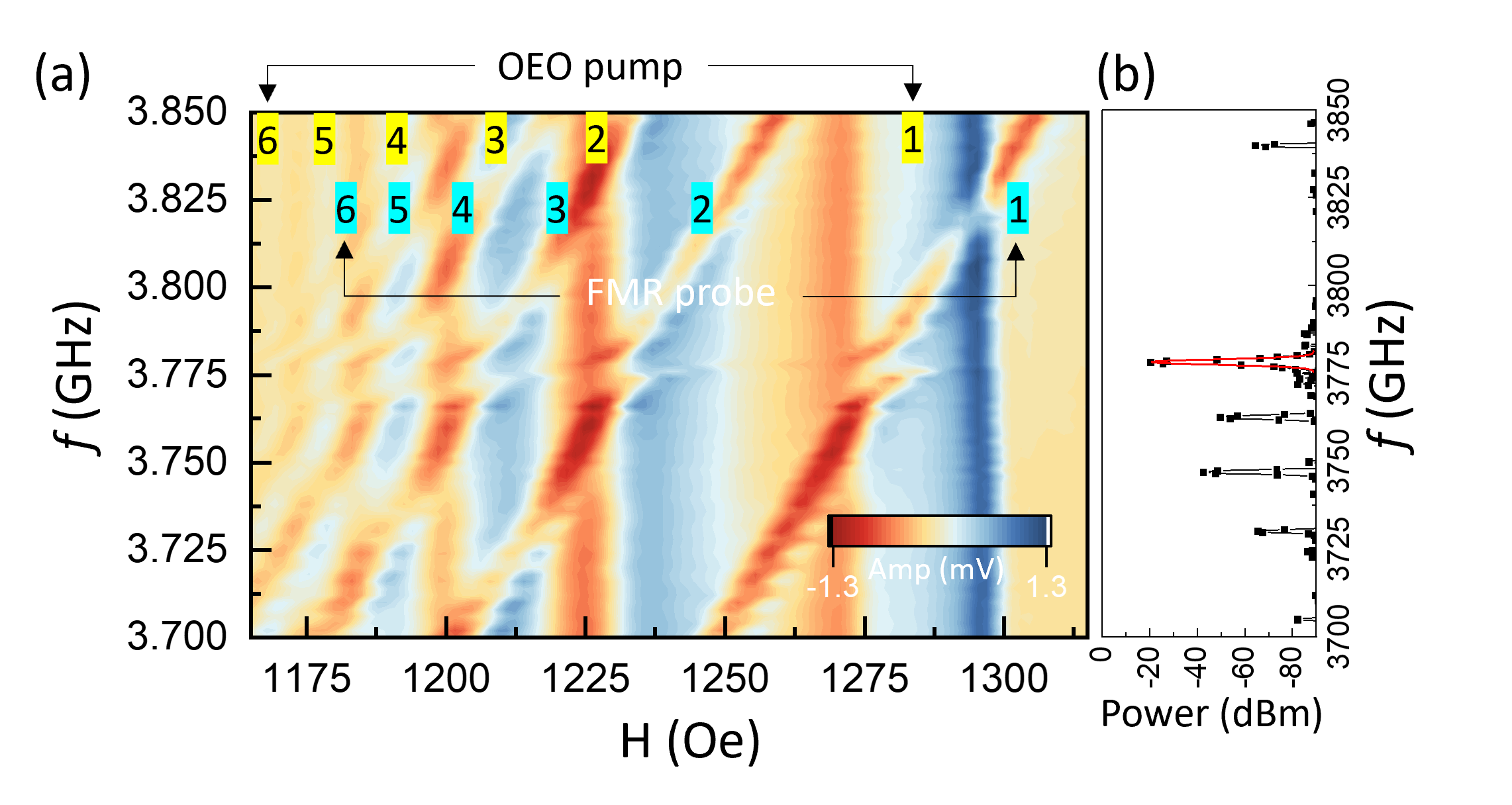}
 \caption{(a) The sub-loop probe signal detected by the lock-in amplifier, showing the coherent coupling between the magnetostatic modes and the OEO pump-induced modes. (b) Power spectrum (from the spectrum analyzer) showing the $2^{nd}$ OEO fundamental mode, resonating $\omega_2 \approx 3778.1$ MHz. \textcolor{black}{A Gaussian fit to this mode yields a photon mode linewidth of 1.53 MHz, which we use in our subsequent theoretical modelling.} All other modes are suppressed by more than 20 dB. }
 \label{fig6}
\end{figure}

In addition, during the process, the strong OEO oscillation signal is also present inside the YIG resonator, \textcolor{black}{acting} as a strong pump to the magnon modes. Such a strong pump leads to a peculiar PIM \cite{pim}, which then causes a pronounced back-action to the probed magnon resonance, depending on the relative detuning of the probe frequency with respect to the OEO oscillation frequency. Figure \ref{fig6}(a) plots the measured lock-in signal as a function of bias magnetic field and probe signal frequency, with the OEO oscillation frequency fixed at ~3778.1 MHz.

We scan the probe signal (low power) near the OEO pump mode  $\omega_2 \approx 3778.1$ MHz, shown in Fig. \ref{fig6}(a). Notably, the pronounced back-action to the probe signal caused by the OEO pump mode leads to a reduction of the magnetostatic magnon amplitude, which, in turn, suppresses the OEO pumping process. The net result is then the observation of anti-crossing features in the $f-H$ dispersion centered around the OEO pump mode, which hybridize with several  magnetostatic modes. \textcolor{black}{These features can modeled as the coherent coupling between the PIM and the associated magnon mode through a Jaynes-Cummings type of interaction of strength $G\approx 11.5$ MHz (from Fig. \ref{fig6}), a function of the magnon occupancy of the mode, which in turn, depends on the intensity of the pump \cite{pim}.}


\subsection{Theory of magnon-induced nonlinear sideband generation and magnon-photon coupling} \label{sec5}
In the absence of YIG, the frequency response of the OEO above the threshold displays equally spaced peaks analogous to that of a Fabry-Perot resonator. Therefore, we model the OEO modes as harmonic oscillators with ladder operators $a_n (a_n^{\dagger})$, $n\in \mathbb{Z}_{> 0}$, possessing resonance frequencies $\omega_n=n\omega_1$, where $\omega_1$ characterizes the frequency of the first fundamental OEO mode. The Hamiltonian of the OEO system is given by
\begin{align}\label{eq5.1}
    H_{oeo}/\hbar=\sum_i\omega_ia_i^\dagger a_i.
\end{align}
The Hamiltonian of the YIG sphere, which accounts for  spin-spin, dipole-dipole and Zeeman interactions, can be diagonalized as \cite{walker,spinw}
\begin{align}\label{eq5.3}
{H}_m/\hbar=\sum_{\textbf{{k}}}\omega_{\textbf{{k}}}m_{\textbf{{k}}}^{\dagger}m_{\textbf{{k}}},
\end{align}
where $m_\textbf{k}$ ($m_\textbf{k}^\dagger$) is the magnon annihilation (creation) operator and $\omega_{\textbf{k}}$  the magnon dispersion, which depends linearly on the applied bias magnetic field $H$. The magnon and the OEO modes interact coherently through the evanescent fields emanating from the waveguide. \textcolor{black}{Furthermore, magnons can couple to the YIG phonon modes (depicted in Fig. \ref{fig6.5}) via the magnetostrictive interaction as discussed earlier.}  

There are several magnon and OEO modes interacting coherently through the evanescent fields. For the sake of brevity, here we focus on the interaction between the second fundamental OEO mode and one of the Walker modes, where the operator $a_2$ ($m_k$) is superseded by $a$ ($m$) with corresponding frequency $\omega_a$ ($\omega_m$). The approximate Hamiltonian of the interacting system, which can be straightforwardly generalized into other modes of the system, is given by
\begin{align}\label{eq5.4}
    H/\hbar=\omega_a a^{\dagger}a+\omega_m m^{\dagger}m+\Omega b^{\dagger}b+g_{am}(a^\dagger m+a m^{\dagger}) \nonumber \\+g_{mb}m^{\dagger}m(b+b^{\dagger})+i\epsilon(a^{\dagger}e^{-i\omega_a t}-ae^{i\omega_a t}).
\end{align}
Here $b$ ($b^{\dagger}$) denotes the annihilation (creation) operator of a phonon mode of frequency $\Omega$, $g_{ma}$ and $g_{mb}$ are respectively, the magnon-photon and magnon-phonon interaction strengths, whereas $\epsilon$ parameterizes the strength of the field pumping the OEO mode. The dynamics of the system in the frame rotating at frequency $\omega_a$ is provided by the following mean-field equations:
\begin{align}\label{eq5.5}
    \dot{a}&=-\kappa a-ig_{am}m+\epsilon, \nonumber\\
    \dot{m}&=-i(\Delta_m-i\gamma_m)m-ig_{am}a-ig_{mb}m(b+b^{\dagger})\nonumber, \\
    \dot{b}&=-i(\Delta_b-i\gamma_b)b-ig_{mb}m^{\dagger}m,
\end{align}
where $\Delta_\alpha=\omega_\alpha-\omega_a$ ($\alpha\in\{m,b\}$) while $\kappa$ and $\gamma_\alpha$ are, respectively the damping rates of the associated modes. In the long-time limit, the general solution to the modes in Eq. (\ref{eq5.5}) takes the form
\begin{align}\label{eq5.6}
    a=\sum_{n\in \mathbb{Z}}\mathcal{A}_n e^{-in\Omega t}, \nonumber \\
    m=\sum_{n\in \mathbb{Z}} \mathcal{M}_n e^{-in\Omega t}, \nonumber \\
    b=\sum_{n\in \mathbb{Z}}\mathcal{B}_n e^{-in\Omega t}.
\end{align}
Equation (\ref{eq5.6}) shows that the nonlinear interactions $\propto g_{mb}$ lead to the generation of sidebands oscillating at frequencies displaced by an integer multiple of $\Omega$ around $\omega_a$. The coefficients $O_j$ ($O\in\{\mathcal{A},\mathcal{M},\mathcal{B}\}$) can be obtained by substituting Eq. (\ref{eq5.6}) into Eq. (\ref{eq5.5}). For instance,
\begin{align}\label{eq5.7}   \mathcal{A}_0=\frac{\epsilon(\gamma_m+i\tilde{\Delta}_m)}{\kappa(\gamma_m+i\tilde{\Delta}_m)+g_{am}^2},
\end{align}

\begin{figure}
 \centering
 \includegraphics[scale=0.6]{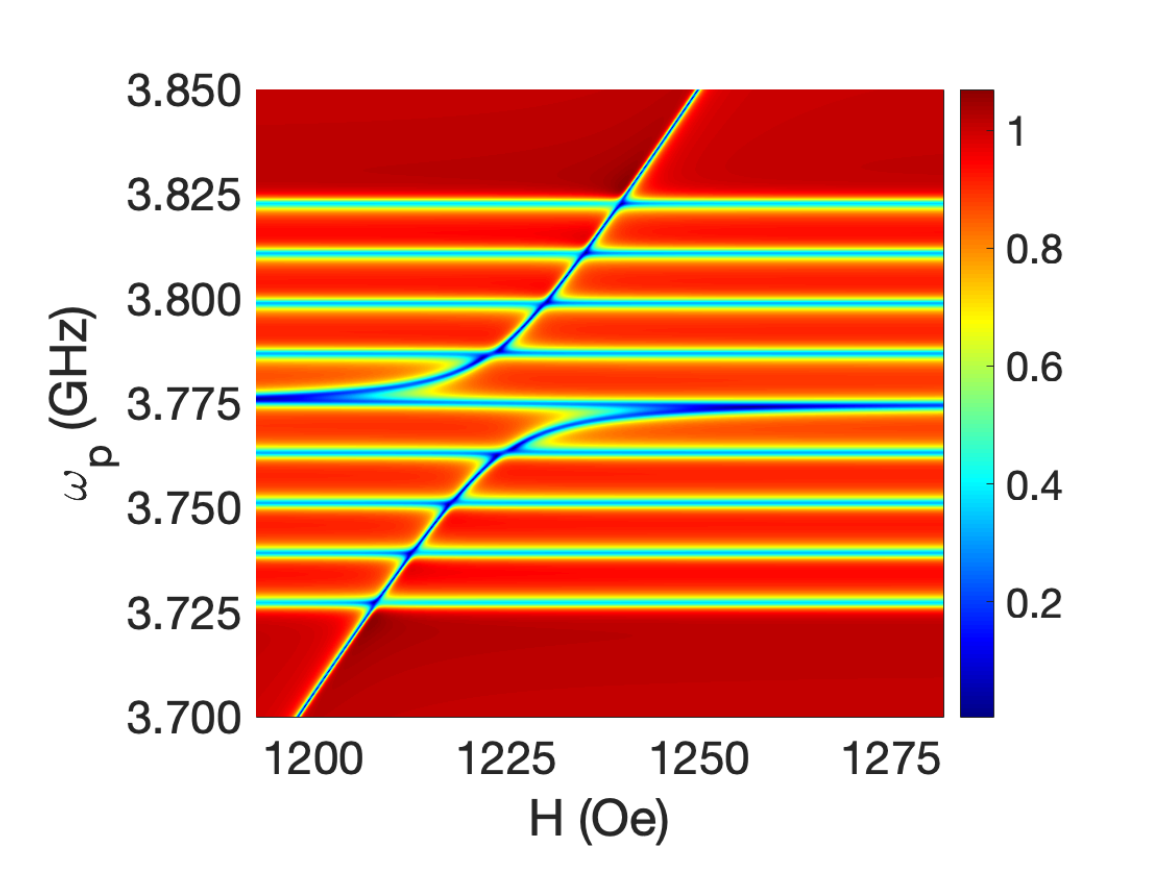}
 \caption{The absolute value of the transmission $|t|$ as a function of the probe frequency $\omega_p$ and bias magnetic field $H$. Numerical value of the other parameters are \textcolor{black}{$\kappa/2\pi=1.53$ MHz, $\gamma/2\pi=2\alpha\omega_m\approx 1$ MHz, and $g_{ma}/\gamma\approx\Gamma/\gamma=1$ and the coupling strength $G/\gamma=11.5$ from Fig. \ref{fig6}.}}
 \label{fig7}
\end{figure}
where $\tilde{\Delta}_m=\Delta_m+\frac{2g_{mb}^2\Delta_b |\mathcal{M}_0|^2}{\Delta_b^2+\gamma_b^2}$. The sidebands get self-sustained through the OEO and the corresponding modes appear in the power spectra as shown in Fig. \ref{fig2}. The approximate Hamiltonian of the second fundamental OEO mode and associated sidebands interacting with the magnon mode can be modelled as
\begin{align}\label{eq5.7}
    H_{eff}/\hbar=\omega_aa^{\dagger}a+\sum_{n\in \mathbb{Z}-0}(\omega_a+n\Omega)a_n^\dagger a_n+\omega_a m_d^\dagger m_d\nonumber \\
    +\omega_m m^\dagger m+\sum_{n\in \mathbb{Z}-0}g_{am}(a_n^{\dagger}m+h.c.)+g_{am}(a^{\dagger}m+h.c.)\nonumber \\
    +G(m^{\dagger}m_d+h.c.),
\end{align}
where $a_n$ ($a_n^{\dagger}$) denotes the annihilation (creation) operator of the $n^{th}$ sideband while $m_d$ is the pump-induced mode. The strength of coupling $G$ between the magnon and the pump-induced mode is a function of the intensity of the incident electromagnetic field. Information regarding the eigenstructure of the system can be extracted from the transmission to a weak probe field of frequency $\omega_p$ applied on the system. The input-output relation relevant to the probe transmission is given by
\begin{align}\label{eq5.8}
    a^{in}-a^{out}=\sqrt{\Gamma}\sum_{n\in \mathbb{Z}-0}(a_n+a+m_d+m),
\end{align}
where $a^{in}$ and $a^{out}$ characterize the input and transmitted fields respectively, while $\Gamma$ is the strength of coupling between the probe field and the modes of the system. In Fig. \ref{fig7}, we plot the numerically calculated transmission coefficient $t={\frac{a^{out}}{{a^{in}}}}$ by keeping eight sidebands of the fundamental OEO modes \footnote{The frequency $\omega_m=\gamma_e (H+H_0)$ where $\gamma_e$ represents the gyromagnetic ratio and $H_0$ is a shift chosen to match one of the Walker modes in Fig. \ref{fig6}.}. \textcolor{black}{We take the rate of coupling between OEO and magnon modes approximately equal to the coupling between the input probe field and the modes of the system, i.e., $g_{am}\approx \Gamma$. It is also worth mentioning that small variations to the coupling strengths do not impact the physics of Fig. \ref{fig7}. Other parameters $\kappa$ and $G$ are respectively obtained from a Gaussian fit and level-splitting in Fig. \ref{fig6}}. Clearly, the level-splitting at the fundamental OEO frequency (due to photon-magnon coupling with the pump-induced mode) is significantly larger than the one displayed by the sidebands (due to phonon-magnon coupling), in agreement with the experimental observations of level-repulsion showcased in Fig. \ref{fig6}. 

In summary, we investigate a new \textcolor{black}{hybrid platform comprising of an optoelectronic oscillator coupled with a YIG sphere integrated on a microwave waveguide. We find that both the OEO loop's power and spectrum are significantly affected by the additional magnon resonator. The presence of strong electromagnetic fields from the OEO excites the magnetostrictive interaction between magnon and phonon modes of the magnetic sphere, leading to the generation of newer frequencies around the fundamental OEO modes displaced by phonon frequency. These newer frequencies become self-sustained through the OEO by the time delay of the system and get manifested as sidebands in both the power spectrum of the OEO and waveguide transmission.}

\textcolor{black}{Due to the system's complexity, a number of characteristic oscillations are observed: (i) the high order OEO modes featuring dense oscillation pattern with 1.55 MHz FSR in Fig. \ref{fig3}(c); (ii) the OEO eigenmodes, i.e. the main, pronounced oscillation shown in Fig. \ref{fig2}(a) and (b), which are frequencies that satisfy the OEO oscillation condition; (iii) additional oscillation series due to the non-flat gain profile, which manifest as sidebands near the main oscillation modes (with a splitting of 18.6 MHz and 17.3 MHz); (iv) phonon modes induced by the magnetostrictive interaction, forming sidebands with a 13 MHz FSR, in Fig. \ref{fig4}; (v) magnetostatic (Walker) modes shown in Fig. \ref{fig6} with mode order (\textit{m}, \textit{m}, 0). }  

Our experimental observations are well-reproduced by a coupled mode theory and numerical simulations. Thanks to the excellent tunability \textcolor{black}{of magnon system} and the \textcolor{black}{``long-range-coherent" OEO system}, our apparatus renders a novel platform to explore both classical and quantum phenomena involving solid-state systems \textcolor{black}{and may potentially be used for achieving remote magnon-photon coupling beyond the near fields.}    

\section*{Data availability}
The datasets used and/or analysed during the current study available from the corresponding author on reasonable request.

\section*{Acknowledgment}
The experimental work at UNC-CH was supported by the U.S. National Science Foundation under Grant No. ECCS-2246254. B. Flebus acknowledges support from National Science Foundation under Grant No. NSF DMR-2144086.

\section*{Competing interests}
All authors declare no financial or non-financial competing interests. 

\section*{Author Contributions}
B.F. and W.Z. conceived the idea. Y.X., A.C.,J.C., and W.Z. performed the experiment and data analysis. J.N. and B.F. developed the theoretical model. A.P. and X. Z. performed the COMSOL simulation. All authors contributed to the writing of the manuscript.


\section*{Figure Legends} 

\textbf{FIGURE 1. Schematic illustration of the setup:} consisting of the (main loop) optoelectronic, magnonic (YIG sphere), and fiber delay parts. Bandpassing functions are given by the enclosed circulator pair and dedicated bandpass filters while other rf components are rated broadband. The fiber splitter splits half of the optical power to a microwave diode for system power monitoring. In between the two circulators, a ferromagnetic resonance probe (sub-loop) is installed that probes the YIG magnonic characteristics at a much lower rf power compared to the main loop (negligible disturbance).

\textbf{FIGURE 2. OEO eigen-modes} (using a short fiber delayline of 100-m) under two different loop power levels, (a) low power: 7 dBm, and (b) high power: 14 dBm, \textcolor{black}{measured by the rf power diode in the microwave branch.} Power spectra with and without the YIG magnonic load are shown. \textcolor{black}{At 7 dBm loop power, the first 3 eigen-modes are observed. At 14 dBm loop power, a total of 9 discernible eigen-modes are observed, labeled 1 to 9. } (c) The extra magnonic-load induces a frequency shift $\Delta f = f_{OEO}-f_{YIG}$, corresponding to a small additional time delay, which is linearly proportional to the \textcolor{black}{OEO harmonics, $f_{OEO}$ (mode index)}.

\textbf{FIGURE 3. Magnon-induced modulation of OEO spectra. }(a) Contour plot of the power spectrum scanned near the FMR regime of a magnon mode driven by the $1^{st}$ OEO photon mode. (b) The FMR signal (probed by the sub-loop at low power) and the power diode signal (main loop) versus the external magnetic field. The power diode measures primarily the power of the OEO main mode, which is at least 20dB higher than all other sidebands at off-resonance. (c) Power spectrum snapshots at selective magnetic fields. Around the magnetic resonance, the OEO's main photon mode is attenuated, whose energy then smears out and distributes to the continuum. \textcolor{black}{The dense oscillations correspond to the free-spectral-range (FSR) of high-order OEO modes, which is 1.55 MHz (645 ns delay) with a 100 m fiber.}

\textbf{FIGURE 4. Emergence of photonic sidebands. } (a) Power spectrum at zero magnetic field showing the main OEO mode and the two pronounced sidebands at a fiber length of 5 m. The left and right sidebands are 13 MHz away from the central band. (b) The scanned [$f,H$] dispersion near the OEO mode by using the probe sub-loop. A series of anticrossing gaps are observed by the hybridization between the mixed photon mode series and the YIG distinct magnetostatic (Walker) modes. 

\textbf{FIGURE 5. Phonon mode simulation.} Simulated total displacement of the phonon mode at 13 MHz on a 1-mm-diameter YIG sphere.

\textbf{FIGURE 6. Magnetostatic modes.} (a) The magnon signal (probed by the sub-loop) and (b) the power diode signal (main loop) versus the external magnetic field. A total of six magnetostatic modes ($m=1,2,3,4,5,6$) of YIG can be observed due to the OEO pump.

\textbf{FIGURE 7. Magnon-photon coupling. } (a) The sub-loop probe signal detected by the lock-in amplifier, showing the coherent coupling between the magnetostatic modes and the OEO pump-induced modes. (b) Power spectrum (from the spectrum analyzer) showing the $2^{nd}$ OEO fundamental mode, resonating $\omega_2 \approx 3778.1$ MHz. \textcolor{black}{A Gaussian fit to this mode yields a photon mode linewidth of 1.53 MHz, which we use in our subsequent theoretical modelling.} All other modes are suppressed by more than 20 dB.

\textbf{FIGURE 8. Theoretical modeling. } The absolute value of the transmission $|t|$ as a function of the probe frequency $\omega_p$ and bias magnetic field $H$. Numerical value of the other parameters are \textcolor{black}{$\kappa/2\pi=1.53$ MHz, $\gamma/2\pi=2\alpha\omega_m\approx 1$ MHz, and $g_{ma}/\gamma\approx\Gamma/\gamma=1$ and the coupling strength $G/\gamma=11.5$ from Fig. \ref{fig6}.}


\begin{thebibliography}{1}%
\makeatletter
\providecommand \@ifxundefined [1]{%
 \@ifx{#1\undefined}
}%
\providecommand \@ifnum [1]{%
 \ifnum #1\expandafter \@firstoftwo
 \else \expandafter \@secondoftwo
 \fi
}%
\providecommand \@ifx [1]{%
 \ifx #1\expandafter \@firstoftwo
 \else \expandafter \@secondoftwo
 \fi
}%
\providecommand \natexlab [1]{#1}%
\providecommand \enquote  [1]{``#1''}%
\providecommand \bibnamefont  [1]{#1}%
\providecommand \bibfnamefont [1]{#1}%
\providecommand \citenamefont [1]{#1}%
\providecommand \href@noop [0]{\@secondoftwo}%
\providecommand \href [0]{\begingroup \@sanitize@url \@href}%
\providecommand \@href[1]{\@@startlink{#1}\@@href}%
\providecommand \@@href[1]{\endgroup#1\@@endlink}%
\providecommand \@sanitize@url [0]{\catcode `\\12\catcode `\$12\catcode `\&12\catcode `\#12\catcode `\^12\catcode `\_12\catcode `\%12\relax}%
\providecommand \@@startlink[1]{}%
\providecommand \@@endlink[0]{}%
\providecommand \url  [0]{\begingroup\@sanitize@url \@url }%
\providecommand \@url [1]{\endgroup\@href {#1}{\urlprefix }}%
\providecommand \urlprefix  [0]{URL }%
\providecommand \Eprint [0]{\href }%
\providecommand \doibase [0]{http://dx.doi.org/}%
\providecommand \selectlanguage [0]{\@gobble}%
\providecommand \bibinfo  [0]{\@secondoftwo}%
\providecommand \bibfield  [0]{\@secondoftwo}%
\providecommand \translation [1]{[#1]}%
\providecommand \BibitemOpen [0]{}%
\providecommand \bibitemStop [0]{}%
\providecommand \bibitemNoStop [0]{.\EOS\space}%
\providecommand \EOS [0]{\spacefactor3000\relax}%
\providecommand \BibitemShut  [1]{\csname bibitem#1\endcsname}%
\let\auto@bib@innerbib\@empty
\bibitem [{Note1()}]{Note1}%
  \BibitemOpen
  \bibinfo {note} {The frequency $\omega _m=\gamma _e (H+H_0)$ where $\gamma _e$ represents the gyromagnetic ratio and $H_0$ is a shift chosen to match one of the Walker modes in Fig. \ref {fig6}.}\BibitemShut {Stop}%
\end{thebibliography}%


\begin{thebibliography}{19}

\bibitem{nakamura_apex2019} D. Lachance-Quirion, Y. Tabuchi, A. Gloppe, K. Usami, and Y. Nakamura, ``Hybrid quantum systems based on magnonics", Applied Physics Express \textbf{12}, 070101 (2019).

\bibitem{hu_ssp2018} M. Harder and C. -M. Hu, ``Cavity Spintronics: An Early Review of Recent Progress in the Study of Magnon-Photon Level Repulsion”, Solid State Physics, \textbf{70}, 47 - 121 (2018). R. Stamps and R. Camley (Ed.), Academic Press.

\bibitem{bhoi_ssp2019} B. Bhoi and S. -K. Kim, ``Photon-magnon coupling: Historical perspective, status, and future directions”, Solid State Physics, \textbf{69}, 1 - 77 (2019). R. Stamps and H. Schultheiss (Ed.), Academic Press.

\bibitem{tqe_2021} D. D. Awschalom, C.H.R. Du, R. He, J. Heremans, A. Hoffmann, J. Hou, H. Kurebayashi, Y. Li, L. Liu, V. Novosad, J. Sklenar, S. Sullivan, D. Sun, H. Tang, V. Tyberkevych, C. Trevillian, A. W. Tsen, L. Weiss, W. Zhang, X. Zhang, L. Zhao, and Ch. W. Zollitsch, ``Quantum Engineering With Hybrid Magnonics Systems and Materials", IEEE Trans. Quantum Engineering \textbf{2}, 5500836 (2021).

\bibitem{jap_2021} Y. Li, W. Zhang, V. Tyberkevych, W.-K. Kwok, A. Hoffmann, V. Novosad, ``Hybrid magnonics: Physics, circuits, and applications for coherent information processing", J. Appl. Phys. \textbf{128}, 130902 (2020).

\bibitem{nakamura_science2015} Y. Tabuchi, S. Ishino, A. Noguchi, T. Ishikawa, R. Yamazaki, K. Usami, and Y. Nakamura, ``Coherent coupling between a ferromagnetic magnon and a superconducting qubit", Science \textbf{349}, 405 (2015).

\bibitem{haigh_prb2019} L. McKenzie-Sell, J. Xie, C.-M. Lee, J. W. A. Robinson, C. Ciccarelli, and J. A. Haigh, ``Low-impedance superconducting microwave resonators for strong coupling to small magnetic mode volumes", Phys. Rev. B \textbf{99}, 140414 (2019).

\bibitem{yili_prl2019} Y. Li, T. Polakovic, Y.-L. Wang, J. Xu, S. Lendinez, Z. Zhang, J. Ding, T. Khaire, H. Saglam, R. Divan, J. Pearson, W. K. Kwok, Z. Xiao, V. Novosad, A. Hoffmann, and W. Zhang, ``Strong Coupling between Magnons and Microwave Photons in On-Chip Ferromagnet-Superconductor Thin-Film Devices", Phys. Rev. Lett. \textbf{123}, 107701 (2019).

\bibitem{luqiao_prl2019} J. T. Hou and L. Liu, ``Strong coupling between microwave photons and nanomagnet magnons", Phys. Rev. Lett. \textbf{123}, 107702 (2019).

\bibitem{huebl_prl2013} H. Huebl, C. W. Zollitsch, J. Lotze, F. Hocke, M. Greifenstein, A. Marx, R. Gross, and S. T. B. Goennenwein, ``High Cooperativity in Coupled Microwave Resonator Ferrimagnetic Insulator Hybrids", Phys. Rev. Lett. \textbf{111}, 127003 (2013).

\bibitem{bai_prl2015} L. Bai, M. Harder, Y. P. Chen, X. Fan, J. Q. Xiao, and C. M. Hu, ``Spin Pumping in Electrodynamically Coupled Magnon-Photon Systems", Phys. Rev. Lett. \textbf{114}, 227201 (2015).

\bibitem{xufeng_prl2014} X. Zhang, C.-L. Zou, L. Jiang, and H. X. Tang, ``Strongly Coupled Magnons and Cavity Microwave Photons", Phys. Rev. Lett. \textbf{113}, 156401 (2014).

\bibitem{nonlinearity_yuan2023} Shasha Zheng, Zhenyu Wang, Yipu Wang, Fengxiao Sun, Qiongyi He, Peng Yan, H. Y. Yuan, ``Tutorial: Nonlinear magnonics", arXiv:2303.16313 (2023). 

\bibitem{bauer_prb2020} Mehrdad Elyasi, Yaroslav M. Blanter, and Gerrit E. W. Bauer, ``Resources of nonlinear cavity magnonics for quantum information", Phys. Rev. B \textbf{101}, 054402 (2020). 

\bibitem{prl_rao2023} Jinwei Rao, C. Y. Wang, Bimu Yao, Z. J. Chen, K. X. Zhao, Wei Lu, ``Meter-scale strong coupling between magnons and photons", Phys. Rev. Lett. in press (2013).

\bibitem{gdp} Bimu Yao, Y.S. Gui, J.W. Rao, Y.H. Zhang, Wei Lu, and C.-M. Hu, ``Coherent Microwave Emission of Gain-Driven Polaritons", Phys. Rev. Lett. \textbf{130}, 146702 (2023).

\bibitem{pim} J.W. Rao, Bimu Yao, C.Y. Wang, C. Zhang, Tao Yu, and Wei Lu, ``Unveiling a Pump-Induced Magnon Mode via Its Strong Interaction with Walker Modes", Phys. Rev. Lett. \textbf{130}, 046705 (2023). 

\bibitem{stancil} Daniel D. Stancil, \textit{Spin Waves: Theory and Applications}, Springer, 2009 Edition. 

\bibitem{magnon_bec} S. O. Demokritov, V. E. Demidov, O. Dzyapko, G. A. Melkov, A. A. Serga, B. Hillebrands, and A. N. Slavin, ``Bose–Einstein condensation of quasi-equilibrium magnons at room temperature under pumping", Nature 443, 430 (2006). 

\bibitem{magnon_bec1} A. J. E. Kreil, D. A. Bozhko, H. Yu. Musiienko-Shmarova, V. I. Vasyuchka, V. S. L’vov, A. Pomyalov, B. Hillebrands, and A. A. Serga, ``From Kinetic Instability to Bose-Einstein Condensation and Magnon Supercurrents", Phys. Rev. Lett. 121, 077203 (2018).

\bibitem{magnon_bec2} M. Schneider, D. Breitbach, R. O. Serha, Q. Wang, A. A. Serga, A. N. Slavin, V. S. Tiberkevich, B. Heinz, B. Lägel, T. Brächer, C. Dubs, S. Knauer, O. V. Dobrovolskiy, P. Pirro, B. Hillebrands, and A. V. Chumak, ``Control of the Bose-Einstein Condensation of Magnons by the Spin Hall Effect", Phys. Rev. Lett. 127, 237203 (2021). 

\bibitem{magnon_bec3} M. Mohseni, A. Qaiumzadeh, A. A. Serga, A. Brataas, B. Hillebrands, and P. Pirro, ``Bose–Einstein condensation of nonequilibrium magnons in confined systems", New J. Phys. 22 083080 (2020). 

\bibitem{helmut_prap2023} T. Hache, L. Körber, T. Hula, K. Lenz, A. Kákay, O. Hellwig, J. Lindner, J. Fassbender, and H. Schultheiss, ``Control of Four-Magnon Scattering by Pure Spin Current in a Magnonic Waveguide", Phys. Rev. Applied \textbf{20}, 014062 (2023). 

\bibitem{you_prl2022} R.-C. Shen, J. Li, Z.-Y. Fan, Y.-P. Wang, and J. Q. You, ``Mechanical Bistability in Kerr-Modified Cavity Magnomechanics", Phys. Rev. Lett. \textbf{129}, 123601 (2022).

\bibitem{yan_prl2021} Z. Wang, H.-Y. Yuan, Y. Cao, Z.-X. Li, R. A. Duine, and P. Yan, ``Magnonic Frequency Comb Through Nonlinear Magnon-Skyrmion Scattering", Phys. Rev. Lett. \textbf{127}, 037202 (2021).

\bibitem{helmut_apl2022} T. Hula, K. Schultheiss, F. J. T. Gonçalves, L. Körber, M. Bejarano, M. Copus, L. Flacke, L. Liensberger, A. Buzdakov, A. Kákay, M. Weiler, R. Camley, J. Fassbender, and H. Schultheiss, ``Spin-wave frequency combs", Appl. Phys. Lett. \textbf{121}, 112404 (2022).

\bibitem{akerman_prb2020} S. Muralidhar, A. A. Awad, A. Alemán, R. Khymyn, M. Dvornik, D. Hanstorp, and J. Åkerman, ``Sustained coherent spin wave emission using frequency combs", Phys. Rev. B \textbf{101}, 224423 (2020).

\bibitem{xiong_review2023} H. Xiong, ``Magnonic frequency combs based on the resonantly enhanced magnetostrictive effect", Fundamental Res. \textbf{3}, 8 (2023).

\bibitem{larger_rmp2019} Yanne K. Chembo, Daniel Brunner, Maxime Jacquot, and Laurent Larger, ``Optoelectronic oscillators with time-delayed feedback", Rev. Mod. Phys. \textbf{91}, 035006 (2019). 

\bibitem{moeo} Yuzan Xiong, Zhizhi Zhang, Yi Li, Mouhamad Hammami, Joseph Sklenar, Laith Alahmed, Peng Li, Thomas Sebastian, Hongwei Qu, Axel Hoffmann, Valentine Novosad, and Wei Zhang, ``Experimental parameters, combined dynamics, and nonlinearity of a magnonic-opto-electronic oscillator (MOEO)", Rev Sci Instrum \textbf{91}, 125105 (2020).  

\bibitem{ustinov_magnlett2015} A. B. Ustinov, A. A. Nikitin, and Boris A. Kalinikos, "Magnetically Tunable Microwave Spin-Wave Photonic Oscillator", IEEE Magn. Lett. \textbf{6}, 3500704 (2015).

\bibitem{vitko_magnlett2018} V. V. Vitko, A. A. Nikitin, A. B. Ustinov,  and Boris A. Kalinikos, "Microwave Bistability in Active Ring Resonators With Dual Spin-Wave and Optical Nonlinearities", IEEE Magn. Lett. \textbf{9}, 3506304 (2018).

\bibitem{watt1} Stuart Watt, Mikhail Kostylev, Alexey B. Ustinov, and Boris A. Kalinikos, ``Implementing a Magnonic Reservoir Computer Model Based on Time-Delay Multiplexing", Phys. Rev. Applied \textbf{15}, 064060 (2021).

\bibitem{watt2} Stuart Watt and Mikhail Kostylev, ``Learning Trajectories from Spin-Wave Dynamics", Phys. Rev. Applied \textbf{19}, 064029 (2023). 

\bibitem{watt3} Stuart Watt and Mikhail Kostylev, ``Reservoir Computing Using a Spin-Wave Delay-Line Active-Ring Resonator Based on Yttrium-Iron-Garnet Film",  Phys. Rev. Applied \textbf{13}, 034057 (2020). 

\bibitem{you_prb2016} Yi-Pu Wang, Guo-Qiang Zhang, Dengke Zhang, Xiao-Qing Luo, Wei Xiong, Shuai-Peng Wang, Tie-Fu Li, C.-M. Hu, and J. Q. You, ``Magnon Kerr effect in a strongly coupled cavity-magnon system", Phys. Rev. B \textbf{94}, 224410 (2016).

\bibitem{zhang_sciadv2016} Xufeng Zhang, Chang-Ling Zou, Liang Jiang, and Hong Tang, ``Cavity magnomechanics", Sci. Adv. \textbf{2}, e1501286 (2016).


\bibitem{walker} L. R. Walker, ``Resonant Modes of Ferromagnetic Spheroids", J. Appl. Phys. \textbf{29}, 318–323 (1958). 

\bibitem{spinw} D. D. Stancil and Anil Prabhakar, ``Spin waves" Vol. \textbf{5}. New York: Springer, (2009).


\end{thebibliography}
\end{document}